# A Markov Chain Approach for Defining the Fundamental Efficiency Limits of Classical and Bifacial Multi-junction Tandem Solar Cells

Muhammad A. Alam[1] and M. Ryyan Khan[1]

[1]School of Electrical and Computer Engineering, Purdue University, West Lafayette, IN, USA

(email: alam@purdue.edu )

*Abstract*—Bifacial tandem cells promise to reduce three fundamental losses (above-bandgap, below bandgap, and the uncollected light between panels) inherent in classical single junction PV systems. The successive filtering of light through the bandgap cascade, and requirement of current continuity make optimization of tandem cells difficult, accessible only to numerical solution through computer modeling. The challenge is even more complicated for bifacial design. In this paper, we use an elegantly simple Markov chain approach to show that the essential physics of optimization is intuitively obvious, and deeply insightful results can obtained analytically with a few lines of algebra. This powerful approach reproduces, as special cases, all the known results of traditional/bifacial tandem cells, and highlights the asymptotic efficiency gain of these technologies.

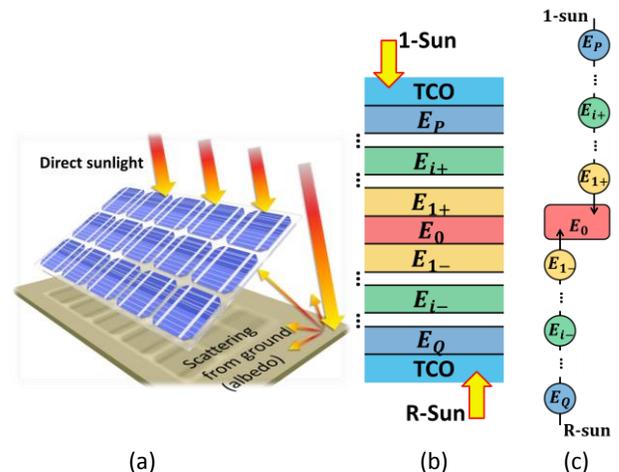

Fig. 1. (a) A bifacial panel collects both the direct sunlight and the light scattered from ground (albedo reflectance, $R$). (b) A bifacial multijunction tandem (B-MJT) is shown. The cell receives 1-Sun and $R$-Sun illumination from the top and back respectively. (c) The B-MJT shown in (b) can be viewed as a bubble (Markov) chain.

## I. INTRODUCTION

The optimum single junction (SJ) solar cell fails to convert 2/3 of the incident sunlight into useful energy [1]. In fact, these unconverted sub-bandgap (sub-BG) and above-bandgap (above-BG) photons further degrade the performance and reliability through self-heating [2], [3]. Moreover, the panels in a solar farm must be spatially separated to avoid shadowing; as a result, ~50% of the photons are wasted in the space in between [4]. With this 'space loss' accounted for, ~83% of the sunlight incident on a solar farm will never be converted to electricity.

A bifacial multi-junction tandem cell (B-MJT) promises to stem these three fundamental losses as follows: photons of various energies are converted by the sequence of absorbers with decreasing bandgaps so that 'sub-BG' and 'above-BG' losses are reduced in half [5]. In addition, bifacial cells partially recover (~30% in practice) the space-loss by converting the albedo light [6]–[9], see Fig. 1(a). Therefore, in principle, the B-MJT solar farm may be 250% more efficient than a SJ farm.

Since the 1960s, many groups have analyzed the physics and optimized the design of MJT with finite number of cells [5], [10], [11]. In contrast, bifacial cells are relatively new, but their high efficiency and reduced temperature sensitivity have sparked commercial interest. The thermodynamics and optimization of two-junction bifacial cells have been reported recently [8], [9]. The results show that the optimization is nontrivial: In a classical MJT, the need for current-matching dictates sequential decrease in bandgap from the top to the bottom. In B-MJT, the bottom cell is illuminated by albedo light, therefore, we need not maintain the bandgap sequence; a partial inversion of bandgaps is possible and desirable.

Even in the idealized thermodynamic limit, however, many questions remain unanswered: What is the optimum bandgap sequence of a 5-junction B-MJT and how does it compare to classical MJT? How would the configuration change when the solar farm is installed on grass vs. concrete rooftop? At what point, does the marginal gain of an additional junction is negligible?

A numerical simulation can answer these questions, but the essential physics are sometimes lost in the fog of numerical modeling. Instead, here we use a simple approximation for BG-dependent photocurrent, within a Markov chain formulation [12], to show that the choice of BG in classical vs. B-MJT is described by an elegantly simple formulation. The optimum efficiency predicted by the simple model match the numerical results within 2%. Away from the optimum BG, the fluorescence coupling is essential and numerical modeling cannot be avoided. Even in those cases, the results of the calculation provide excellent initial guesses regarding the potential of the bifacial cell technology.

## II. THE MARKOV CHAIN MODELING FRAMEWORK

Fig. 1(b) represents the typical configuration of a bifacial cell. Conceptually, a B-MJT may be represented, as in Fig. 1(c), by a chain of bubbles (each representing a material with bandgap, $E_g$, and short-circuit current, $J_{sc}(E_g)$), illuminated by 1-sun on the top and $R$-sun at the bottom. The cell with the smallest bandgap ($E_0 \equiv E_{g,min}$) is located at $\{0\}$. The chain-segment illuminated by the direct incident light is marked $\{i+\}$ with the top cell at $\{P\}$. Similarly, the cells illuminated by the albedo light is marked $\{i-\}$, and the bottom cell is $\{Q\}$. Thus, the total number of cells is $N \equiv P + Q + 1$.

Assuming complete absorption above the bandgap, the current in the individual bubbles is related to short circuit current, $J_{sc}(E_g)$ of isolated absorbers, as follows:

$$J_{\{i,\pm\}} = J_{sc,i\pm} - J_{sc,(i+1)\pm}, \text{ except that} \quad (1a)$$

$$J_P = J_{sc,P}, \quad (1b)$$

$$J_Q = R J_{sc,Q}. \quad (1c)$$

Since the current through the series connected cells must be identical, the equations above are numerically equal.

Despite the complexity of the AM1.5G spectrum (or AM1.5D, AM0 for that matter), the short-circuit current, $J_{sc}(E_G)$, scales almost linearly within the bandgap range ($0.5\ eV < E_G < 2.0eV$), i.e.,

$$J_{sc}(E_G) = J_0[1 - \beta E_G] \quad (2)$$

where $\beta$ is a constant, and $J_0$ depends on intensity, $I$.

Inserting, (2) into (1) and dictating that the current must be continuous through the Markov chain, we find that the bandgap optimization problem fully solved simply as follows,

$$[E] = [M]^{-1}[Z] \quad (3)$$

where $[E] = [E_P, ... E_{i,+}.. E_{j,-}, ... , E_Q]$ is the bandgap vector of size $N - 1$ (excluding $E_0$), and the residual vector, $[Z]$, of the same size is given by

$$[Z] = \begin{cases} -[1, 0, 0, ..... \beta(1+R)E_0, \beta(1+R)E_0, ... 0, 0, R]; & P, Q > 0 \\ -[1, 0, 0, ..... \beta(1+R)E_0 - R]; & P > 0, Q = 0 \\ -[1 + \beta(1+R)E_0 - R]; & P = 1, Q = 0 \end{cases}$$

and, $[M] \equiv \begin{bmatrix} \beta \nabla_P^2 & B_R \\ B & R\beta \nabla_Q^2 \end{bmatrix}$

where, $R$ is the effective albedo reflectance and,

$$[\nabla^2] \equiv \begin{bmatrix} -2 & 1 & \cdots & \cdots \\ 1 & -2 & 1 & \cdots \\ \cdots & \cdots & \cdots & \cdots \\ \cdots & \cdots & 1 & -2 \end{bmatrix}, [B] \equiv \begin{bmatrix} 0 & \cdots & -\beta \\ \vdots & \ddots & \vdots \\ 0 & \cdots & 0 \end{bmatrix}, [B_R] \equiv \begin{bmatrix} 0 & \cdots & 0 \\ \vdots & \ddots & \vdots \\ -\beta R & \cdots & 0 \end{bmatrix}$$

Note that, $[\nabla_P^2]$ is a $P \times P$ matrix. Once vector $[E]$ is specified, the full $J - V$ characteristics

$$J(V) = J_{ph} - J_{dark}(V) \quad (4)$$

can be determined as follows, see Fig. 2. The photocurrents are matched, therefore, can be replaced by a single source, evaluated $J_{ph} = J_P$ for example. And, the dark current is

$$J_{dark}(V) \equiv J_{D,i} = A_i(E_i)\left(e^{\frac{qV_i}{kT}} - 1\right) \quad (4a)$$

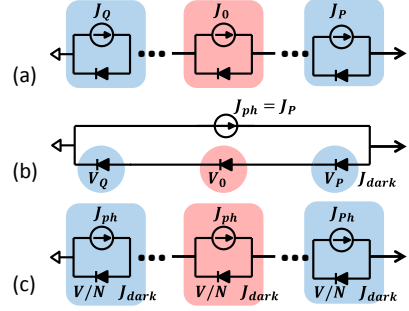

Fig. 2. Circuit model for analyzing the B-MJT.

Here, $A_i(E_g) = 2\pi q\, \gamma(E_g,T) e^{-\frac{E_g}{kT}}$, and $\gamma(E_g,T) \equiv \left(\frac{2kT}{c^2h^3}\right)(E_g^2 + 2kT^2E_g + 2k^2T^2)$ [1]. Using Eq. (4b), we find:

$$V = \sum_N V_i = \sum_N \left(\frac{kT}{q}\right)\ln\left(\left(\frac{J_{dark}}{A_i}\right)+1\right) = \frac{kT}{q}\ln\frac{J_{dark}^N}{\prod A_i}$$

$$\therefore J_{dark}(V) \approx q\, 2\pi\, \gamma_i e^{-\left(\frac{\langle E_g \rangle}{kT}\right)} e^{\frac{qV}{NkT}} \quad (4b)$$

Here, $\langle E_g \rangle$ and $\{E_g\}$ are the arithmetic and geometric means, respectively, of $[E]$, obtained from Eq. (3). And, $\gamma_i$ is the geometric mean of $[\gamma(E_i)]$. *In this remarkable result, Eq. 4(b) suggests that the terminal response of complex B-MJT can be represented by a string of identical cells repeated N-times, making the vast literature of SJ physics available to MJT analysis.*

To summarize, once $E_0, N, P$, and $R$ are specified, Eq. (3) provides the analytical solution of the bandgaps, so that Eq. (1) and Eq. (4) can be used to construct the $J - V$ characteristics and the efficiency, $\eta_T^*(E_0, N, P, R)$, of the cells. To calculate the $P_{max}$, first we find from Eq. (4) that

$$\frac{qV_{opt}}{N} \approx \left(\langle E_g \rangle \left(1 - \left(\frac{T_D}{\langle E_g \rangle}\right)\left(\frac{E_{g,P}}{T_S}\right)\right) - k_B T_D \ln\left(\frac{\Omega_D}{\Omega_S}\right)\right) \quad (5)$$

so that $P_{max} \equiv J(V_{opt})V_{opt}$ gives optimum output. A basic scientific calculator can be used to solve the optimization problem in just a few minutes. Eq. (5) reduces the well-known SJ formula with $\langle E_g \rangle = E_{g,P}$, as expected.

The limitations of the model are evident in the derivation: current is presumed linear in bandgap; the current should be matched at the maximum power-point, not at short-circuit; and it neglects the thermal resistance that self-heats the stack and changes the bandgap. Regardless, the accuracy of the model can be verified against the numerical model, as follows.

III. AN ILLUSTRATIVE EXAMPLE

*A. Bandgap Sequence.*

Let us consider a special case when $Q = 0$ as an illustration of the power of the technique. For arbitrary $P$ and $R$, we have $[Z] \equiv -[1, 0, ..., \beta(1+R)E_0 - R]$. Eq. (3) is now easily solved:

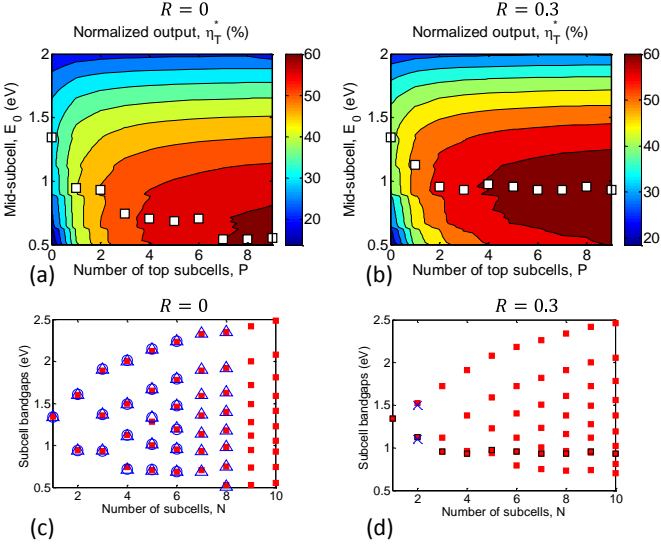

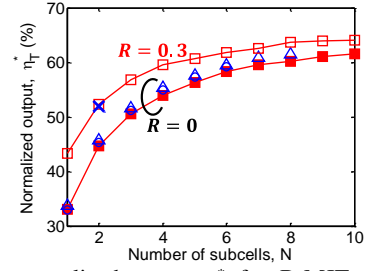

Fig. 4. The normalized output $\eta_T^*$ for B-MJT are shown for $R = 0$ and $R = 0.3$ by filled and open squares. Results are compared to literature: $\Delta$, O, $\times$ are from [9]–[11].

Fig. 3. The normalized B-MJT output $\eta_T^*$ is for $Q = 0$ found as functions of $P$ and $E_0$ at (a) $R = 0$ and (b) $R = 0.3$. The optimum $E_0$ is marked as white squares. (c) and (d) shows the corresponding optimum B-MJT bandgaps (red-filled squares). Results are compared to literature: $\Delta$, O, $\times$ are from [9]–[11].

$$E_i = \left(\frac{i}{\beta N}\right) + \frac{\{(N-i)[\beta(1+R)E_0 - R]\}}{\beta N}. \quad (6)$$

With $R = 0$, the equation reduces to the conventional tandem structure. The reduction in the optimized bandgap for B-MJT, compared to classical MJT, by $(\Delta E_i = -(N-i)(1-\beta E_0)]R/\beta N$ reflects that the fact, with that the bottom is no longer dependent on the filtered light through the top cell for its current; therefore, improved current matching is possible even with reduced bandgap difference. As an aside, when $E_i > 2.0$ eV, the $J_{sc} - E_g$ is no longer linear, $E_i$ from Eq. (6) should be trivially replaced by the $E_i^*$, where $J(E_i) = J(E_i^*)$.

### B. Thermodynamic Limit.

We now calculate, based on Eq. (4) and algorithm discussed in Sec. II, the $\eta_T^*(E_0, N, Q = 0, R)$ and for $N = 1 \ldots 10$ and $R = 0, 0.3$, and plot the results in Figs. 3(a, b), respectively. For comparison, $\eta_T^*$ is the output normalized to 1-sun input. The white squares mark optimum $E_0^{opt}(N)$ that maximizes $\eta_T^*$ for a specified number of junctions. Figs. 3(c) and (d) show that the $E_i$ associated with $E_0^{opt}$ is near perfect agreement with results reported in the literature for the classical and bifacial cells. Given this level of agreement of the bandgaps shown in Fig. 3, it is not surprising that $\eta_T^*$ matches as well, see Fig. 4.

Indeed, for $Q = 0$, the $E_0(N, R)$ optimum is easily derived:

$$E_0^{opt} \approx \left(E_{SJ} - \left(\frac{(N-1)(1-R)}{2\beta N}\right)\right) \frac{2N}{(N(1+R) + (1-R))}. \quad (7)$$

Here, $E_{SJ} = 1.33$ eV is the SJ optimum bandgap. Eq. (7) anticipates asymptotic limit of $E_0^{opt}(N \to \infty)$, see Fig. 3 (c,d).

### IV. CONCLUSIONS

While the results for $R = 0$ (classical tandem) is only of pedagogical interest, the results shown in Fig. 4 is the **first report** of efficiency gain of B-MJT with $N \geq 3$. The results suggest that a 4-junction B-MJT would outperform a 7-junction classical MJT, such is the power of the current-constraint relaxed by the bifacial concept. For the same $N$, the increased power-input of B-MJT would make the cells slightly hotter, but the reduced temperature coefficient of some of the bifacial cells, such as HIT, would compensate the effect. Finally, for $N = 10$ and $R = 0.3$, the gain advantage saturates to approximately 2.5%, a small but significant increase. This however is not the limit: The $\eta_T^*$ would improve further for $Q > 0$ configurations, especially for $N > 4$. The results will be reported in the conference.

To conclude, we have developed a new methodology that can be used to answer broad range of questions regarding tandem cells. At extremely small and very large bandgaps, or for optimization at the maximum power point involving luminescent coupling, numerical simulation would still be necessary, and the final design must rely on careful optimization of finite absorption, reflection, series resistance. Regardless, the methodology reported here stands out in its simplicity and versatility to quantitatively predict a range of phenomena previously accessible only to numerical modeling.

*We acknowledge Prof. J. L. Gray and Prof. P. Bermel for helpful discussions. This work was made possible by financial support of the DOE-SERIIUS and NSF- NEEDS centers.*

### REFERENCES

[1] L. C. Hirst and N. J. Ekins-Daukes, "Fundamental losses in solar cells," *Prog. Photovolt: Res. Appl.*, vol. 19, no. 3, pp. 286–293, 2011.
[2] S. Dongaonkar, C. Deline, and M. A. Alam, "Performance and Reliability Implications of Two-Dimensional Shading in Monolithic Thin-Film Photovoltaic Modules," *IEEE Journal of Photovoltaics*, vol. 3, no. 4, pp. 1367–1375, Oct. 2013.
[3] T. J. Silverman, M. G. Deceglie, X. Sun, R. L. Garris, M. A. Alam, C. Deline, and S. Kurtz, "Thermal and Electrical Effects of Partial Shade in Monolithic Thin-Film Photovoltaic Modules," *IEEE Journal of Photovoltaics*, vol. 5, no. 6, pp. 1742–1747, Nov. 2015.
[4] A. Luque and S. Hegedus, Eds., *Handbook of Photovoltaic Science and Engineering, Second Edition*. 2011.
[5] C. H. Henry, "Limiting efficiencies of ideal single and multiple energy gap terrestrial solar cells," *J. Appl. Phys.*, vol. 51, no. 8, pp. 4494–4500, Aug. 1980.
[6] U. A. Yusufoglu, T. M. Pletzer, L. J. Koduvelikulathu, C. Comparotto, R. Kopecek, and H. Kurz, "Analysis of the Annual Performance of Bifacial Modules and Optimization Methods," *IEEE Journal of Photovoltaics*, vol. 5, no. 1, pp. 320–328, Jan. 2015.
[7] A. Luque, E. Lorenzo, G. Sala, and S. López-Romero, "Diffusing reflectors for bifacial photovoltaic panels," *Solar Cells*, vol. 13, no. 3, pp. 277–292, Jan. 1985.

[8] R. Asadpour, R. V. K. Chavali, M. R. Khan, and M. A. Alam, "Bifacial Si heterojunction-perovskite organic-inorganic tandem to produce highly efficient (ηT* ~ 33%) solar cell," *Applied Physics Letters*, vol. 106, no. 24, p. 243902, Jun. 2015.

[9] M. R. Khan and M. A. Alam, "Thermodynamic limit of bifacial double-junction tandem solar cells," *Applied Physics Letters*, vol. 107, no. 22, p. 223502, Nov. 2015.

[10] S. P. Bremner, M. Y. Levy, and C. B. Honsberg, "Analysis of tandem solar cell efficiencies under AM1.5G spectrum using a rapid flux calculation method," *Prog. Photovolt: Res. Appl.*, vol. 16, no. 3, pp. 225–233, May 2008.

[11] A. S. Brown and M. A. Green, "Detailed balance limit for the series constrained two terminal tandem solar cell," *Physica E: Low-dimensional Systems and Nanostructures*, vol. 14, no. 1–2, pp. 96–100, Apr. 2002.

[12] M. A. Alam and R. K. Smith, "A phenomenological theory of correlated multiple soft-breakdown events in ultra-thin gate dielectrics," in *Reliability Physics Symposium Proceedings, 2003. 41st Annual. 2003 IEEE International*, 2003, pp. 406–411.